# Holographic purification complexity

Elena Cáceres, Josiah Couch, Stefan Eccles, and Willy Fischler

*Theory Group, Department of Physics, The University of Texas at Austin, Austin, Texas 78712, USA*

We study holographic subregion complexity, and its possible connection to purification complexity suggested recently by Agón *et al.* In particular, we study the conjecture that subregion complexity is the purification complexity by considering holographic purifications of a holographic mixed state. We argue that these include states with any amount of coarse-graining consistent with being a purification of the mixed state in question, corresponding holographically to different choices of the cutoff surface. We find that within the complexity = volume and complexity = spacetime volume conjectures, the subregion complexity is equal to the holographic purification complexity. For complexity = action (CA), the subregion complexity seems to provide an upper bound on the holographic purification complexity, though we show cases where this bound is not saturated. One such example is provided by black holes with a large genus behind the horizon, which were studied by Fu *et al.* As such, one must conclude that these offending geometries are not holographic, that CA must be modified, or else that holographic subregion complexity in CA is not dual to the purification complexity of the corresponding reduced state.

## I. INTRODUCTION

In the past several years, quantum complexity has entered into discussions of quantum gravity and holography, starting with a discussion of complexity and the firewall paradox in [1] and later in [2]. Motivated by these considerations, it was later suggested that in the program of bulk reconstruction, boundary data about entanglement is insufficient to reconstruct a dual geometry and that perhaps something like the quantum circuit complexity of the state could fill the gaps [3]. The complexity $C(|\Psi\rangle)$ of a state $|\Psi\rangle$ is the minimum number of basic unitaries, or "gates" $g$, drawn from a preestablished gate set $G = \{g_i\}$, needed to build a circuit $Q = \prod_{i=1}^C g_i$ such that $d_f(Q|\Psi_0\rangle, |\Psi\rangle) < \epsilon$, where $|\Psi_0\rangle$ is a preestablished "reference state," $\epsilon$ is the "tolerance parameter," and $d_f(\cdot, \cdot)$ is the Fubini-Study metric. Out of these ideas arose the "complexity = volume" conjecture [3,4] which speculates that the volume of a maximal spatial slice is dual to the quantum circuit complexity of the dual quantum state living on the intersection of that spatial slice with the boundary. For aesthetic reasons, it was eventually suggested to replace complexity = volume (henceforth referred to simply as CV) with "complexity = action" (CA) [5,6], and while the bulk of the discussion on the topic is concerned with these two conjectures, at the boundaries at least two other speculations exist, namely "complexity = spacetime volume" (CV2.0) [7] and CA-2 [8].[1] CA modifies CV by replacing the volume of a maximal spatial slice by the action evaluated on the Wheeler-DeWitt (WDW) patch associated with the slice, i.e., the causal development of the (UV-regulated[2]) slice. CV2.0 modifies CA by replacing the action on the WDW patch with the spacetime volume of the WDW patch.

All of these conjectures could be applied to any geometry, including some subregion of the bulk, though we would not expect these quantities computed for a general bulk subregion to have any particular meaning. However, considering that the state on a subregion of the boundary is dual to the entanglement wedge, as argued in [9–11], it is tempting to say that these proposals applied to an entanglement wedge might be dual to the complexity of the reduced state on the corresponding boundary subregion. This has been suggested by a number of authors [12,13], and the "complexity" thus computed is termed "subregion complexity." This, however, raises the question: What is the complexity of a mixed state? There is not a unique way to extend the usual definition of circuit complexity from pure states to mixed states, so which extension are we talking about? To answer this question, the authors of [14] considered a number of possible definitions of mixed state complexity and compared them to what happens in holography. They came to the conclusion that the "purification complexity," which is roughly defined as the minimum state complexity among pure states which reduce to the appropriate density matrix on a subsystem, is a good candidate to be dual to subregion complexity in CA. On the other hand, none of the definitions they considered provides a likely dual to subregion complexity in CV. In this paper we further investigate purification complexity, first as defined in [14] but also with minor variations, providing an independent discussion of its behavior for general quantum states and specifically motivating its connection to holographic subregion complexity.

---

[1]Though in general distinct, CA-2 reduces to CV 2.0 for Einstein-Hilbert gravity with no sources other than a cosmological constant, and as such we will not consider CA-2 separately in this paper.

[2]For a slice not cut off by some UV regulator, the WDW patch only corresponds to the causal development of the slice if one forgets the reflecting boundary conditions at infinity.

The purification complexity of a holographic mixed state can be bounded from above by considering all its holographic purifications. Beginning with the entanglement wedge dual to the mixed state in question, all holographic geometries which geodesically complete the wedge provide a family of purifications in their boundary dual states. In any such geometry, we argue (supported by an analogy to the entanglement of purification) that different choices of the cutoff surface in the region complement to the original entanglement wedge correspond to different coarse-grainings of the purifying state. By minimization over cutoffs we argue on general grounds that the purification complexity is bounded above by the subregion complexity in all of CV, CA, and CV2.0. We then prove that so long as we restrict our attention to holographic purifications, this inequality is saturated in CV and CV2.0, because complexity in these proposals is superadditive.

The situation for CA, by contrast, is more complicated. In the absence of a superadditivity property, we must worry that the subregion complexity is not truly minimal among all holographic purifications. To examine this possibility, we consider geodesic completions of the one-sided Bañados-Teitelboim-Zanelli (BTZ) geometry dual to a thermal state. We find that there are indeed cases where the true minimum is smaller than the subregion complexity, thereby contradicting the conjecture in question.

Our results thus differ here from [14]: we are led to the conclusion that in CV and CV2.0, subregion complexity = purification complexity, whereas in CA it seems either this is not the case, or else that certain asymptotically AdS geometries are not holographic.

The paper is organized as follows: In Sec. II we define purification complexity and investigate its additivity properties on subsystems of general quantum states. We use this analysis to sharpen our expectations on the behavior of any bulk holographic dual to purification complexity. In Sec. III we motivate the connection between purification complexity and holographic subregion complexity by analogy with the concept of "entanglement of purification," of which we give a brief overview. We note that for any geodesic completion of the entanglement wedge dual to our mixed state, there will be one purification which corresponds to a cutoff skirting just outside the entanglement wedge along the Hubeny-Rangamani-Takayanagi (HRT) surface. This purification will have a complexity equal (up to a possible boundary like term discussed at the end of the section) to the subregion complexity. In Sec. IV we prove that complexity according to either CV (in its usual form, without a boundary term) or CV2.0 is superadditive. This, in turn, implies that the complexity of the state dual to any geodesic completion of our entanglement wedge, with any choice of cutoff, must be larger than the subregion complexity as ordinarily defined. We are thus led to the conclusion that in CV and CV2.0 the purification found in Sec. III was indeed optimal among holographic purifications, so in these prescriptions the subregion complexity is a purification complexity.

In Sec. V we consider the case of CA, where the purification found in Sec. III need not be optimal. We consider several families of geodesic completions of the one-sided BTZ geometry. One such family is the $n$ sided genus $g$ generalizations of the two-sided BTZ geometry, where we borrow computations done in [15]. These solutions lead to challenges to the purification complexity interpretation of subregion action, namely that the complexity of certain purifications can be lower than the subregion complexity of the BTZ thermal state, and can even be computed to be negative in some cases, although the issue of the negativity of the action was already raised by [15]. We come to the conclusion that if one is not to abandon the proposal that subregion complexity is dual to purification complexity in CA, one must impose even stricter limits on the geometries considered. We further find that one must impose a limit on the cutoffs considered.

## II. PURIFICATION COMPLEXITY: QUANTUM EXPECTATIONS

In this section, we will explore some aspects of purification complexity as defined in [14]. We first define the quantity as well as discuss ambiguities and variations on the definition which could lead to qualitatively different behavior on subsystems. We then discuss the expected behavior of purification complexity on subsystems. It should be noted that Agón et al. give a compelling but inconclusive argument that purification complexity should be subadditive for the left and right factors of the thermofield double state, and plausibly more generally. We here give an independent discussion indicating that purification complexity is neither superadditive nor subadditive in general. We then place this discussion in the context of the decomposition into basis and spectrum complexities utilized by [14], and discuss how this breakdown is sometimes insufficient for discerning additivity properties. We then conclude this section with a discussion of how these expectations ought to manifest in holographic states, listing some basic consistency checks which must be obeyed by any bulk quantity dual to purification complexity.

### A. Definition and variations

Purification complexity $C^P(\rho)$ of a density matrix $\rho$ is defined in [14] to be the minimum pure state complexity over all its purifications, subject to the constraint that every additional qubit of the purifying system ends up entangled with the original "physical" qubits. That is,

$$C^P(\rho) = \min_{|\psi\rangle \in \mathcal{P}} C(|\psi\rangle) \quad (1)$$

where $\mathcal{P}$ is the set of all purifications $|\psi\rangle$ of $\rho$ which have no separable factors which are also purifications, i.e., there is no decomposition $|\psi\rangle = |\psi_1\rangle \otimes |\psi_2\rangle$ such that $|\psi_1\rangle$ also purifies $\rho$. This last condition guarantees that purification complexity reduces to ordinary state complexity on pure states. If $|\phi\rangle$ is a pure state, then any "purification" of it results in a separable state $|\psi\rangle = |\phi\rangle \otimes |\psi'\rangle$, but since $|\phi\rangle$ is pure and hence a purification of itself, no $|\psi\rangle$ with nontrivial $|\psi'\rangle$ is in our set $\mathcal{P}$, and hence $C^P(\rho_\phi) = C(|\phi\rangle)$ (where $\rho_\phi = |\phi\rangle\langle\phi|$). If we however included such purifications in our minimization, we can at best say that $C^P(\rho_\phi) \leq C(|\phi\rangle)$. In purifications satisfying this criterion, we say that the state on the ancillary Hilbert space is "fully entangled" with the original state $|\psi\rangle$, and we will refer to such purifications as "valid" purifications.

It is easy to imagine alternatives to the above definition which share the same spirit as minimization over purifications, but restricting by more or less the allowed class of purifications. At one extreme we could consider "unrestricted" purification complexity, with the minimum taken over all possible purifications. Such a procedure will not reproduce the usual definition of complexity of pure states, but it does provide a different, competing definition, which in principle could be the one relevant for holography (though we are not making that claim here). On the other hand we could place more stringent conditions on the purifying states, or instead, constrain the ancillary Hilbert space used to purify. As an example of the latter type of constraint, we could dictate that only a Hilbert space of minimal dimension may be used, this being fixed by the rank of the density matrix in question. This last possibility is also compatible with the usual pure state definition, and in fact, discussions of Sec. III indicate that such a restriction may be relevant for subregion complexity.

All of these definitions implicitly assume that a notion of pure state complexity has been defined, not only for the original Hilbert space but for every allowed dimension and form of the purifying Hilbert space. A reference state and gate set must be chosen which scale unambiguously with these Hilbert spaces. Mixed state complexity thus inherits all the same ambiguities as any pure state complexity, and in a sense even more. Though at first disconcerting, this feature is perhaps appropriate considering the holographic conjectures; these presumably employ some natural reference and gate set, each suitably adjustable to any cutoff scale.[3]

For developing intuition with N-qubit systems, one plausible procedure is to take the "all zeros" state as the reference, regardless of N. For the gate set, one could specify a universal gate set on two-qubit systems and then allow the same logical operations on any pair of qubits. This is referred to as a two-local gate set. With any k-local gate set (defined analogously), the same prescription scales unambiguously as the Hilbert space adds or eliminates d.o.f., though the rate at which new gates proliferate with additional d.o.f. depends on the specifics of these choices. This leads only to the restriction that subsystems should not be considered below k qubits.

We hold such prescriptions loosely in mind, but with the exception of a few comments, we will henceforth remain agnostic about the choice of gate set and reference state. Instead, we seek to identify features of purification complexity which transcend these ambiguities and therefore inform our expectations about any holographic dual ever before such specifics are understood.

### B. Additivity properties

First, we focus on states which are fully entangled between two subsystems $A$ and $B$.[4] For these states, purification complexity can easily be seen to obey

$$\mathcal{C}(\rho_A) \leq \mathcal{C}(\rho_{AB}) \quad \text{(fully entangled subsystems)} \quad (2)$$

by noting that any valid purification of $AB$ is also a valid purification of $A$ (or $B$), and so the minimum complexity over valid purifications of $A$ (or $B$) is upper bounded by that of $AB$. This inequality immediately leads to another which we dub "weak superadditivity":

$$\mathcal{C}_A + \mathcal{C}_B \leq 2\mathcal{C}_{AB} \quad \text{(fully entangled subsystems)} \quad (3)$$

where we introduce the notation of using a subscript to denote the subsystem, e.g., $C^P(\rho_A) = C^P_A$.

The proof given above for weak superadditivity breaks down when we consider states which factorize on any subsystem of $A$ or $B$, owing to the constraint that in valid purifications the ancilla system must end up fully entangled with the system it purifies. For example, consider a separable pure state on $AB$. While such a state is undoubtedly a purification of the states on $A$ and $B$ respectively, it is not a valid purification.

---

[3]The need to define mixed state complexity has been most keenly appreciated with the intent to generalize the holographic prescriptions to subsystems, but even considering CFT states at finite cutoff, having traced out some UV degrees of freedom (d.o.f.) one might wonder if even the original holographic complexity conjectures already require a notion of mixed state complexity to be well defined. However, at leading order in $1/N$, total states have no entanglement entropy even at finite cutoff and can be thought of as a pure state living on the reduced Hilbert space. In this paper, we restrict ourselves to this limit, where "total state" entails "pure state."

[4]Here again, by "fully entangled" we mean that no subsystem of either $A$ or $B$ factorizes from the full system, i.e., $A$ is fully entangled with $B$ and $B$ is fully entangled with $A$.

In fact, for pure states which are factorizable on any number of subsystems, purification complexity is demonstrably subadditive over these separable factors. To see this, note that the complexity of each subsystem is an ordinary state complexity (because the state on each subsystem is pure), obtained using gates which act unitarily within that subsystem. The circuits which are individually optimal for these subsystems may also be used together to prepare the full product state, but for that purpose, it may not be optimal since circuits over the whole system may additionally utilize gates which couple the subsystems.[5] This composite circuit's state complexity is the sum of the individual state complexities, and it puts an upper bound on the total state complexity. A nearly identical proof guarantees subadditivity for factorizable systems, regardless of whether or not they are pure:

$$\mathcal{C}_A + \mathcal{C}_B \geq \mathcal{C}_{AB} \quad \text{(factorizable subsystems)} \quad (4)$$

Note that the inequalities (3) and (4) do not contradict each other. In fact it is conceivable that together they bound the span of purification complexity on general subsystems:

$$C_{AB} \leq C_A + C_B \leq 2C_{AB} \quad \text{(not proven)}. \quad (5)$$

However neither of these inequalities is proven in general, rather each is proven for a different corner of state space (factorizable subsystems and completely non-factorizable subsystems, respectively). It is natural to ask whether either of these classes can violate the opposing inequality and whether intermediate classes of states obey either inequality. We return to these question shortly, but pause here to relate these statements to the work of [14] and the decomposition of purification complexity into spectrum and basis components.

### C. Basis and spectrum decomposition

Given an arbitrary mixed state and a large enough ancillary Hilbert space (e.g., a duplicate Hilbert space is always sufficient), it is always possible to construct a purification through a two-part process: first prepare a state with the appropriate spectrum on the reduced system, then from this state rearrange the subsystem basis until the target density matrix is achieved. There is an optimal circuit which performs each of these tasks, and their complexities define the spectrum complexity $C^S$ and the basis complexity $C^B$ respectively.[6] In sequence these operations prepare the full mixed state; there may be more efficient routes to prepare a purification, but the sum of these circuit complexities upper bounds the purification complexity:

$$C^P \leq C^S + C^B. \quad (6)$$

In [14], the authors consider the possibility that any one of these complexities ($C^P$, $C^S$, or $C^B$) might correspond to holographic subregion complexities as computed using the complexity = action (CA) or complexity = volume (CV) conjectures. The best tentative match aligned the CA subregion prescription with the full purification complexity $C^P$. This correspondence was particularly encouraged by the expectation that $C^P$ should be subadditive, and among holographic prescriptions only CA includes bulk contributions which are not positive definite, allowing that at least in the case considered CA was also subadditive. We will revisit this particular holographic example in Sec. V, but we here give a schematic outline of the reason for these expectations.

Consider a two-sided eternal black hole with the subregions being the full left and right boundaries (we will use subscripts L and R for "left" and "right" on a Penrose diagram, and subscript T for "total" or "thermofield double state"). The individual subregions each decompose into a basis and spectrum part,

$$C_L^P \leq C^S + C_L^B,$$
$$C_R^P \leq C^S + C_R^B. \quad (7)$$

Because the combined state is pure, the spectrum part is the same for left and right subsystems. To prepare the total state, we can presumably borrow the circuits utilized in the above decompositions, but importantly the spectrum part need only be prepared once at the beginning of this process:

$$C_T^P \leq C^S + C_L^B + C_R^B. \quad (8)$$

Each of the preceding circuit decompositions only upper bounds the true purification complexity, but if we blithely suppose that the bounds are approximately saturated, then comparing (7) and (8) leads to the expectation that

$$C_T^P \leq C_L^P + C_R^P \leq 2C_T^P. \tag{9}$$

---

[5]It may first seem that if we start with an unentangled reference state, then gates which couple unentangled subsystems should play no role in the optimal circuit because these gates create entanglement. However, though such gates are necessary to create entanglement between the two subsystems, they do not always do so. It is easy to find factorizable states and gate sets where the optimal circuit utilizes these gates without ever creating entanglement between the subsystems, even at intermediate stages.

[6]In [14] what we call basis complexity is denoted $\tilde{C}^B$, while $C^B$ denotes the exact difference $C^P - C^S$. We avoid using this exact difference to ensure that $C^S$ and $C^B$ are the complexities of circuits which can be applied in succession to prepare the correct mixed state.

These inequalities match those of Eq. (5). In the case considered here, with the subsystems being left and right halves of an eternal black hole, the rightmost inequality of (9) follows rigorously from (3) for fully entangled subsystems. The leftmost inequality is less certain. Particularly, in equation (8), we assume the total state can be prepared by "borrowing the circuits" used to prepare the subsystem density matrices. However, it is only guaranteed that this combined circuit will prepare a state with the correct subsystem density matrices. There are many such states, and these can have vastly different complexities; preparing the correct total state may require complex operations which are effectively unnoticed by either subsystem and are not accounted for in any of the $C^S$, $C_L^B$, or $C_R^B$ circuits of Eq. (7). To illustrate this point, recall the behavior of the thermofield double state evolved away from the $t_L = t_R = 0$ slice:

$$U(t_L, t_R)|TFD\rangle = e^{-iE_n(t_L+t_R)}e^{-\beta E_n/2}/\sqrt{Z}|n\rangle_L|n\rangle_R \tag{10}$$

where $U(t_L, t_R) = U_L(t_L) \otimes U_R(t_R) = e^{-iH_L t_L} \otimes e^{-iH_R t_R}$ implements time evolution independently on each boundary (with times set to increase "upward" on both sides of a Penrose diagram). The effects of this operation, apparent in the leading phase factor of (10), go unnoticed by either subsystem and of course, it is precisely these effects which lead to the famed late-time linear growth of the pure state complexity. Away from the time symmetric slice of the two-sided black hole, the growth of the total state complexity will inevitably break the left-hand side inequality in (9).

If we consider the $t_L = t_R = 0$ boundary state only, the total state complexity is minimal among the family of states in (10), and so expected to obey $C_T \leq C_L^P + C_R^P$. Indeed this is the chief expectation which found a match in the complexity = action subregion calculations of [14]. However, there is a subtlety related to the degeneracy of the energy spectrum which may muddle even this expectation. It was pointed out in [16] that when a subsystem density matrix has a degenerate spectrum, it has interesting implications for purification complexity. Unitaries which enact rotations or phase shifts within such a degenerate subspace act trivially on the subsystem density matrix (effectively limiting the basis complexity of the density matrix), so they never contribute to the purification complexity. However the same unitaries can affect the pure state on the combined system, sometimes increasing the complexity of the target state.[7]

---

[7]The unitaries enacting time evolution on the thermofield double state are a special case among this class of unitaries. Even if the spectrum is entirely nondegenerate, there are phase rotations within each energy subspace which go unnoticed by either subsystem density matrix but contribute nontrivially to the total state complexity.

If we consider the most extreme case of a fully degenerate spectrum (or the $T \to \infty$ limit), we have maximally mixed subsystems. Preparing both subsystems is equivalent to preparing N bell pairs. This can be thought of as minimizing complexity over a huge family of states which all prepare the appropriate subsystem density matrices:

$$|\psi\rangle = U_L \otimes U_R \left(\frac{1}{\sqrt{N}} \sum_{i=1}^N |i\rangle_L |i\rangle_R\right). \tag{11}$$

Any state among these is a valid purification, and the subsystem purification complexities are upper bounded by minimizing state complexity overall $U_L$ and $U_R$. The total state, on the other hand, will be some particular state among (11) with particular $U_L$ and $U_R$. If the total state happens to be the minimally complex state among these, then $C_L^P = C_R^P = C_T$ and weak superadditivity is saturated. On the other hand, by choosing $U_R$, $U_L$ to make the total state maximally complex we can engineer the total state to violate subadditivity.

If we consider again the thermofield double state at finite temperature on a time-symmetric slice, the total state complexity is at least minimal among

$$|\psi\rangle_T = e^{-iH_L t_L} \otimes e^{-iH_R t_R} \left(\frac{1}{\sqrt{N}} \sum_{i=1}^N |i\rangle_L |i\rangle_R\right). \tag{12}$$

However, if there is some degeneracy in the energy spectrum, then in preparing the subsystem density matrices we can minimize over a larger class of unitaries (all those which do not intermix degenerate subspaces, but not all $U_A$ and $U_B$). This class of unitaries is still potentially much larger than minimization over the time evolution unitaries which occurs at the $t_L = t_R = 0$ slice. It is not immediately clear how much the "extra minimization" over a larger class of unitaries affects the subsystem complexity in comparison to the total state complexity in the case of the thermofield double state. It would seem to depend on the specifics of the energy eigenstates and the gate set involved.

Unfortunately, through these considerations, we are only able to cast doubt on the expectation that the $t_L = t_R = 0$ thermofield double state ought to have purification complexity which is subadditive on the left and right factors, and not provide a rigorous alternative. We will nevertheless consider this example (specifically the BTZ case) holographically in Sec. V and subject it to other consistency checks.

### D. Expectations for holographic states

The additivity relationships discussed in Sec. II B hold for different types of quantum states. Under what circumstances are these relationships relevant to holographic quantum states? We here consider this question, and then give a short list of consistency checks which must be satisfied by any holographic dual to purification complexity.

A pure state which factorizes between subsystems has no entanglement between those subsystems. On the other hand, the smoothness and connectedness of a holographic spacetime indicate that the state on any boundary subregion is fully entangled with its complement. We, therefore, consider the smoothness and connectedness of the classical geometry as the necessary and sufficient condition preventing factorization of the boundary pure state under geometric partitions. Holographic states dual to connected spacetimes are therefore of the "fully entangled" type (see Sec. II B) under such partitions.

Though connected geometries are dual to fully entangled states, we may still consider factorizable holographic pure states. To do so we merely treat multiple independent holographic geometries as product states, living in separate Hilbert spaces with boundary theories decoupled. This thought experiment provides at least one valuable lesson. Any of the geometric prescriptions for subregion complexity will be exactly additive on such factorizable subsystems. While utilizing gates which couple the combined Hilbert space could in principle allow for increased efficiency in preparing the combined state, evidently the notion of complexity dual to bulk action or volume does not take advantage of such gates. Either holographic states are always of the sort that they are never optimally constructed utilizing these gates, or such gates should be excluded from the outset. This latter possibility stands in tension with naive prescriptions for choosing the gate set to vary only with the size of the Hilbert space with no other concern for locality or the entanglement structure of the state (such as the k-local gate set prescription outlined in Sec. II A). This also indicates that if we consider only holographic purifications, the validity constraint excluding unentangled factors is redundant, because evidently including such factors never results in a "speed up."

Turning again to connected holographic spacetimes, if we consider only such systems we can apply our expectations about fully entangled states (see Sec. II B). Along with the basic requirement that purification complexity is positive definite, we have a primitive list of consistency checks to perform on any proposed holographic dual to purification complexity. For any boundary subregion $A$ and neighboring (connected, but not necessarily small) extension of that boundary subregion $\delta A$, we expect

$$\text{positivity}: C_A^P > 0,$$
$$\text{monotonicity}: C_{A+\delta A}^P > C_A^P,$$
$$\text{weak superadditivity}: C_A^P + C_{\delta A}^P < 2C_{A+\delta A}^P.$$

The monotonicity property, which is symmetric on $A$ and $\delta A$, implies weak superadditivity. The same applies to the case already discussed in Sec. II C, the two sides of an eternal black hole solution. In this case $\delta A$ is replaced with $A^c$, the boundary region complement to $A$. This is not a "neighboring" boundary subregion (they are only connected through the bulk), but they form a partitioning of the total boundary ($T$) and the subsystems are fully entangled,

$$\text{weak superadditivity}: C_A^P + C_{A^c}^P < 2C_T.$$

## III. HOLOGRAPHIC PURIFICATION COMPLEXITY

In this section, we consider in detail the conjecture(s) that volume or action on a subsystem entanglement wedge in holographic geometries might be dual to the purification complexity of the corresponding boundary mixed state. We first motivate the relationship to purification complexity, as opposed to some other notion of mixed state complexity, by considering what the total state complexity conjecture may already imply about the meaning of subregion complexity. We do so through an analogy with the concept of "entanglement of purification," explained in Sec. III A.

In Sec. III B we make the case that the volume prescription for subregion complexity is computing a type of purification complexity. To be precise, if the class of purifications considered is all holographic purifications, then the prescription computes precisely the minimum complexity among these. Over any less restricted class of purifications, the subregion prescription merely bounds from above the purification complexity so defined. The same arguments applied to the action prescription do not lead so inexorably to the notion of purification complexity. They imply that the prescription computes the complexity of a particular purification, but it is not clear that it is the optimal one. We defer more explicit holographic tests of this idea for subregion action to Sec. V.

The considerations of this section imply that the volume prescription matches our general expectations for purification complexity more or less automatically, but in Sec. III C we discuss a puzzle with this interpretation and use it to motivate a modification of the bulk volume prescription to include a boundary term on the HRT surface.

### A. Motivation from entanglement of purification

We would like to test the conjecture that subregion complexity is dual to purification complexity by finding a holographic estimate of the purification complexity (independent of the usual subregion complexity prescription). For guidance on how to construct such a holographic estimate, we turn to discussions *entanglement of purification* in holography. Entanglement of purification is defined as follows: Given a bipartite system consisting of subsystems $A$ and $B$, and a state $\rho_{AB}$ on that system, the entanglement of purification between $A$ and $B$ of $\rho_{AB}$ is the minimum entanglement entropy $S(\rho_{A\bar{A}})$ of the reduced state $\rho_{A\bar{A}}$ minimized over all purifications $|\psi\rangle_{A\bar{A}B\bar{B}}$ of $\rho_{AB}$ and all partitions of the purifying system into $\bar{A}$ and $\bar{B}$. If $\rho_{AB}$ is pure, then the entanglement of purification is simply the entanglement entropy of the reduced state $\rho_A$ and $\rho_B$. Recent work [17–19], has discussed a conjectured holographic dual to this quantity for holographic CFTs, which is given as follows: Consider subregions $A$ and $B$ on the boundary, and the joint entanglement wedge of $AB$. The holographic entanglement of purification is then given by the area of the minimal surface in the entanglement wedge which divides $A$ from $B$ (refer to Fig. 1). In the case where there is no mutual information between $A$ and $B$, and so the entanglement wedge is disjoint, we have that $A$ and $B$ are already divided, and so the entanglement of purification is zero. In the case of two disjoint intervals with nonzero mutual information, the minimal surface goes between the two disconnected pieces of the RT surface, and we need only minimize over which points on the RT surface it will intersect.

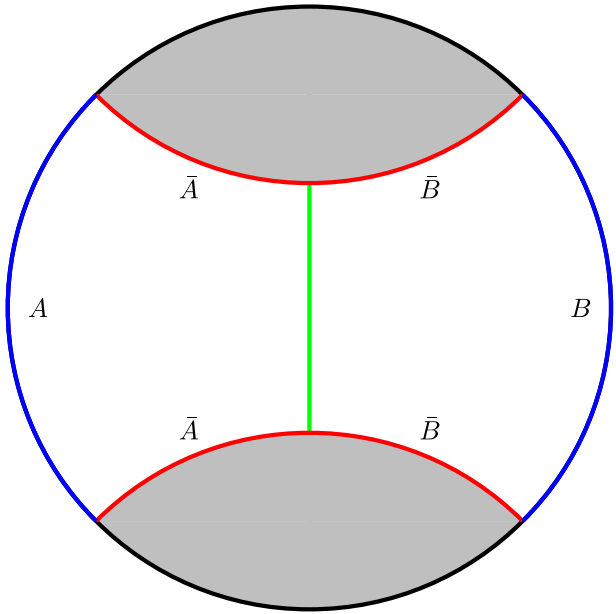

FIG. 1. Entanglement of purification between two interval subregions with nonzero mutual information. The minimal surface whose area gives the entanglement of purification is shown in green.

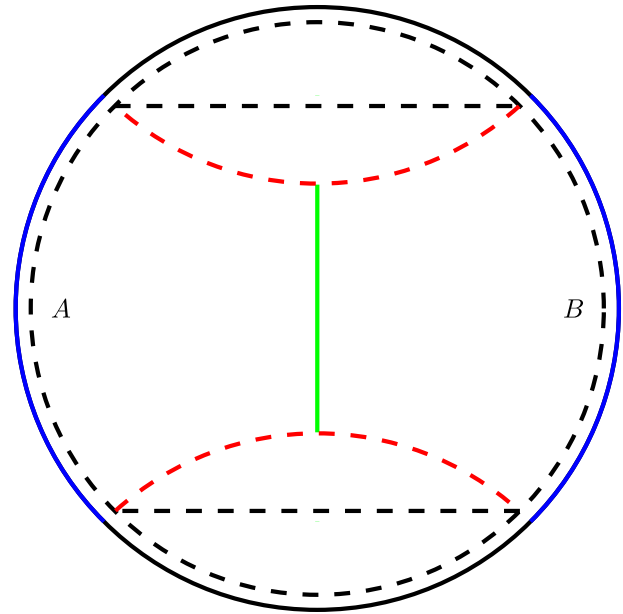

FIG. 2. We may purify the state on $AB$ to states with different course grainings, corresponding to different cutoffs, shown here as a dashed line. The optimal purification will correspond to cutoff which hugs the RT surface, shown here as a red dashed line.

The intuition behind this definition is as follows: There is an optimal purification, which lives on the subregion $AB$ on the boundary, as well as the on the RT surface of $AB$. The auxiliary states on the RT surface can be partitioned into an $\bar{A}$ system and $\bar{B}$ system in different ways, and for each partition, we may compute the entropy of $A\bar{A}$ using the RT prescription. By minimizing over such partitions, we arrive at the entanglement of purification, as given above. But how did we arrive at the fact that the purifying auxiliar system "lives" on the RT surface? This is supported by intuition from tensor networks, but we can justify it as follows:

In minimizing over purifications, we certainly know that the state on the full boundary is a purification of the state on $AB$. However, this state contains much more information than is needed to purify the reduced state $\rho_{AB}$ on $AB$. As a consequence, for any partition, there more entropy than we needed. We can fix this by renormalizing the state, thereby getting rid of the extra information. These renormalized states correspond to putting different cutoffs in the bulk, which agree with our original cutoff in the entanglement wedge of $AB$, but which outside that wedge can be different. Each of these cutoffs corresponds to a different renormalization of the global state, and as such a different purification of $\rho_{AB}$ (see Fig. 2). Clearly then, the cutoff which will give the smallest RT surface, and likewise the one that corresponds to course graining away all the information not needed to purify $\rho_{AB}$, is the one which hugs the RT surface of $AB$. We then get the prescription above.

Of course, there are potentially other holographic purifications of $\rho_{AB}$. Suppose there is an isometric embedding of the entanglement wedge of $AB$ into a geodesically complete asymptotically AdS spacetime, such that the entanglement wedge of the image $\tilde{A}\tilde{B}$ of $AB$ is the image of the entanglement wedge of $AB$. Then according to subregion duality, the reduced state on $\tilde{A}\tilde{B}$ simply is $\rho_{AB}$, since both states are dual to geometries which are identical up to isometry. We may then repeat the above procedure on the new geometry. For entanglement of purification, we will inevitably be led to the same purification, by the arguments above. We will nevertheless term such geometries "purifying geometries" for the holographic state $\rho_{AB}$.

We claim that this same procedure of minimizing over purifications by considering all purifying geometries, and all cutoffs on these purifying geometries compatible with our cutoff in the entanglement wedge, should be applied to find the purification complexity of a given reduced state holographically. We flesh out this claim in the following subsection.

### B. Purification complexity

Suppose we take for granted the original complexity = volume conjecture, that the maximal volume slice asymptoting to a fixed boundary Cauchy slice is dual to some notion of complexity for the total state on that boundary slice.[8] Consider all possible holographic geometries which geodesically complete a fixed entanglement wedge. The states dual to these purifying geometries provide a set of purifications of the density matrix on the original boundary subregion (any pure state is, of course, a purification of all its subsystems). Further consider all possible cutoff surfaces in the complementary portion of these spacetimes, subject to the restriction that they can at least sustain a Cauchy slice for the original entanglement wedge. The corresponding set of coarse-grained states provides an even larger class of purifications to consider.

Trusting the total state complexity conjecture, we can compute all the complexities of these purifications and find the minimum. With the complexity = volume prescription, we are inevitably led to the conclusion that the minimum pure state complexity among all holographic purifications corresponds to choosing a cutoff surface which traces just outside (see next section) the HRT surface for the subregion[9]; deviation from this choice results in either an increase in volume and a higher complexity, or in the exclusion of some portion of the entanglement wedge, which invalidates the state as a purification of the original density matrix. The first statement follows intuitively from the positive definiteness of volume, and more rigorously from the superadditivity relationship proven in the next section. The second statement relies on the specific duality between an entanglement wedge and the corresponding boundary mixed state. Any cutoff contour which cannot sustain a Cauchy slice for the original entanglement wedge will inevitably exclude some bulk operators which ought to be described by the original boundary subregion's mixed state. In so far as we consider only holographic purifications of a boundary mixed state, these considerations imply that the volume subregion prescription on an entanglement wedge is computing precisely the minimum complexity among purifications. The same claims follow analogously for the complexity = spacetime volume prescription.

Now consider the same series of statements for the complexity = action subregion prescription. We can consider purifying geometries of a fixed entanglement wedge, and we can vary over cutoff surfaces. The action on the entanglement wedge submanifold is just another pure state complexity, with the complementary d.o.f. course-grained to the limiting case of the HRT surface itself. However, lacking in these considerations is the idea that computing this particular purification's complexity amounts to a minimization over holographic purifications. The action is not a positive-definite quantity, and other choices of cutoff surfaces in some purifying geometry may provide a purification of lower complexity. Finding any such state amounts to a disproof that the action subregion prescription can be called a purification complexity, in the usual sense of a minimization over the complexity of purifications. We find such counterexamples in Sec. V.

We now provide an example of the procedure just outlined by considering the two-sided black hole (see Fig. 3). If our aim is to estimate the purification complexity of the left boundary mixed state with a certain cutoff $r = \delta_L$, and we are allowed to minimize over the purifying right cutoff, $\delta_R$, an interesting result emerges: When one sets $\delta_R = r_+$, and regularizes the WDW patch by setting its "corners" on the cutoff surfaces, one recovers (in this limit) the usual subregion complexity of the left side. Analogous procedures result in similar conclusions for other two-sided geometries, or for subregions of $AdS_3$. In fact, for a general boundary subregion $A$, allowing a minimization over possible cutoffs in the complement region, subject to certain consistency conditions (e.g., the total cutoff must be continuous), will lead to the inequality

$$C^P(A) \leq C^{\text{subregion}}(A) \qquad (13)$$

whenever the entanglement wedges of a region and of its compliment meet on a single surface. By the HRT

---

[8]In the complexity = volume conjecture, a choice of length scale is necessary to turn a proportionality into a true equality. Traditionally this length scale has simply been taken to be the AdS scale, but recent work [20] has suggested that a variable scale determined by features of the total geometry may be more appropriate. It is not clear how such a variable length scale should be set for the arbitrary subregions considered in this paper. For simplicity, we first consider the case where the length scale is the AdS scale. For our conclusions about additivity to hold for complexity and not just volume in a proposal with variable length scale, we at least require that the length scale is the same for the subregion, the total state, and the complement subregion. In a hypothetical scheme where the length scale changes even with the cutoff surface, the minimization procedures described in this section applies to complexity as well as volume only if variations in the cutoff surface which increase the max volume obey $\delta \log(V/V_0) \geq \delta \log(l/l_0)$, with $l$ being the variable length scale, and $V_0$, $l_0$ being some reference values.

[9]Interestingly, the same holographic purification which "lives on" the HRT surface of a boundary subsystem was also recently identified in [21] as the purification which simultaneously minimizes the entanglement of purification for all bipartitionings of the boundary mixed state.

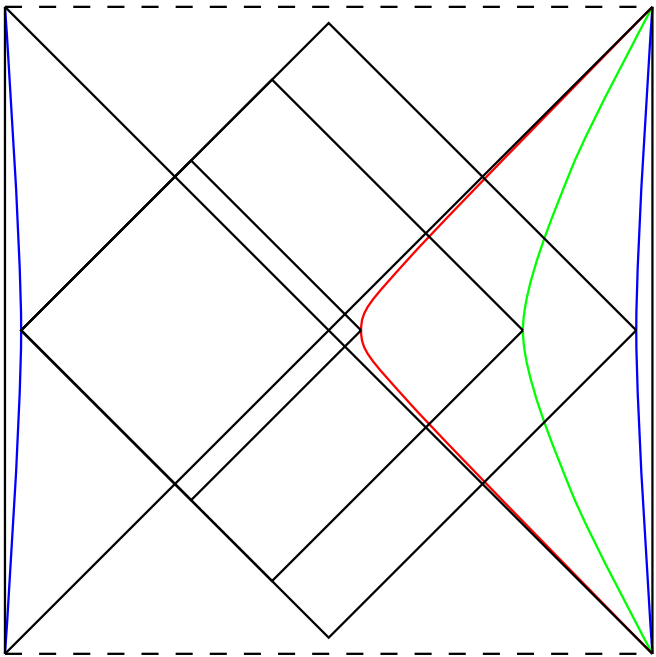

FIG. 3. A two-sided black hole, with different cutoff surfaces on the right side. For each cutoff surface, we have drawn the corresponding regulated WDW patch.

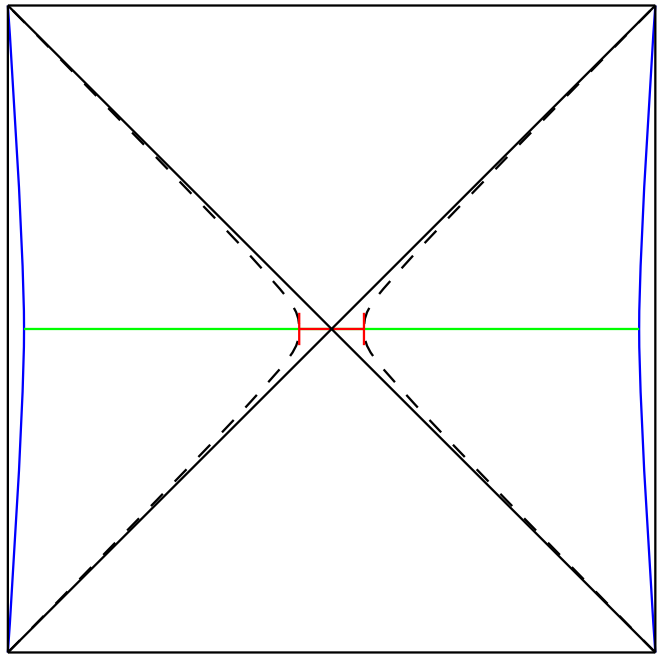

FIG. 4. The dashed lines represent the stretched horizon. The purification complexity in CV of the left (right) reduced state is given by the volume of the left (right) green segment plus the red segment. The complexity of the total state is given by the volume of both green segments plus the volume of the red segment. We thus have that $C_L + C_R - C_T$ is given by the volume of the red segment, and as such is positive definite. We may identify the green segments with the basis complexity, and the red segment with the spectrum complexity, as per the discussion in Sec. 3.3 of [14].

prescription, this should always happen for geodesically complete geometries.

Under either CV or CV2.0, superadditivity holds for any values of the left and right cutoff (we give a proof in Sec. IV), and we thus have that

$$C(|\psi\rangle) \geq C^{\text{subregion}}(A) \qquad (14)$$

for any holographic purification $|\psi\rangle$ of $\rho_A$. However, then this means that if we define a purification complexity only with respect to holographic states (which is perhaps appropriate at leading order in $\frac{1}{N}$), then we have that subregion complexity = holographic purification complexity. This strongly suggests that the duality proposed in [14] for CA works for both CV and CV2.0. The situation with CA is actually less clear, as the action is not positive definite and Eq. (14) does not apply. The inequality (13) does still hold for CA though, provided this minimization over the right cutoff is valid.

### C. Adding a boundary term

One issue with this proposal to minimize over all cutoffs is that, for the optimal purification, the subregion complexity of the reduced state on the purifying right system vanishes. However, this reduced state must have a fixed entanglement entropy, as the state on the whole system is pure. Either we must have a mixed reference state (which seems unusual), or something else must be going on. We can resolve this issue by not allowing the cutoff surface to be pushed all the way to the horizon surface, instead taking it only to the stretched horizon. This too has a number of interesting consequences, the first of which is that it makes the purification complexity in CV subadditive on a slice about which there is time reflection symmetry. This is because the left and right purification complexities both include the volume of the slice between the left and right stretched horizon, and so this part of the volume gets counted twice when one adds the two purification complexities, and only once when one computes the total complexity [see Fig. 4]. This is reminiscent of the argument in Sec. 3.3 of [14] that the spectrum complexity gets double counted when adding the purification complexities, suggesting that perhaps this segment of the volume between the left and right stretched horizons should be identified with the spectrum complexity of the mixed state. We would then identify the part of the volume between the left cutoff and the left stretched horizon as the basis complexity.

When considering the thermofield double state away from the $t_L = t_R = 0$ slice, the maximal slice will not pass through the HRT surface, and the simple "double counting" mentioned in the above case does not apply. However, for these total states, we argued (see Sec. II C) that the spectrum/basis decomposition of the subsystems is insufficient to parse total state complexity. In these cases, the behind-the-horizon region probed by the maximal slices should roughly correspond to operators which mix degenerate subspaces and go unnoticed by either subsystem density matrix. See [22] for a recently proposed definition of a related quantity, the "binding complexity."

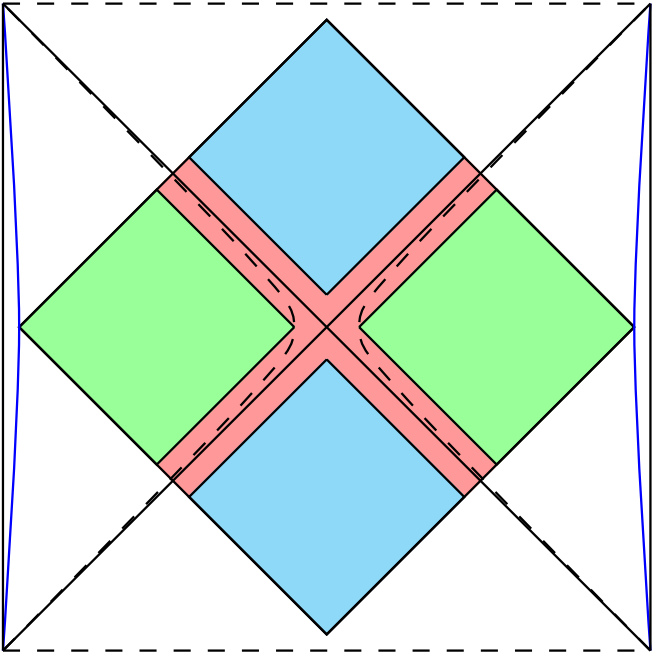

FIG. 5. Similarly to CV, we may decompose the action in CA, CV2.0, or CA-2 by associating different parts of the WDW patch to the spectrum, basis, and degeneracy complexity. Roughly speaking, we suggest the green regions should correspond to the basis complexity, the blue regions to the degeneracy complexity, and the red to the spectrum complexity.

At this stage, one might be worried that we have ruined the upper bound of purification complexity by subregion complexity. However, our bound on purification complexity differs from subregion complexity by a term proportional to the area of the HRT surface (namely the volume between the true horizon and the stretched horizon), which could thus be thought of as an extra boundary term in the subregion complexity prescription. We will take the view that such a discrepancy likely should be corrected by altering the definition of subregion complexity in this way.[10]

Such boundary terms could also be included for CV2.0, CA, and CA-2. In these cases, however, one does not automatically get subadditivity. For CV2.0 in particular, there will generally be a significant contribution to the total state complexity from behind the entanglement horizon which cannot be identified with either the spectrum or basis complexity (see Fig. 5). As suggested above for CV off of the $t=0$ slice, this can perhaps be identified with the complexity in the total state due to gates acting within a degenerate eigenspace of the left and right density matrices, or even rotating the relative phase for a nondegenerate eigenspace. Discussions of this sort have occurred in [23], with proposals for decomposing a holographic spacetime into subregions of particular significance for the quantity of "uncomplexity" (though those proposals did not strictly include a boundary term).

---

[10]We would like to thank Phuc Nguyen for suggesting including a boundary term in CV.

## IV. SUPERADDITIVITY

It was noted in [14] that in CV, holographic subregion complexity obeys

$$C(\rho_{AB}) \geq C(\rho_A) + C(\rho_B), \qquad (15)$$

when $A$ and $B$ partition a complete boundary Cauchy slice. Recently, a related property for "uncomplexity" was discussed in [16] for quantum systems. In this section we provide an independent discussion of this property, which we will call "superadditivity." We will give a proof that subregion complexity obeys this property for general subregions in CV, as well as in CV2.0, and discuss whether it may hold in CA.

Note that superadditivity implies that the subregion complexity of a given subregion must be less than or equal to the complexity of any of its holographic purifications. This will hold regardless of the cutoff imposed, as it is merely a statement about maximal volumes on Lorentzian manifolds with boundary. If we suppose (as is reasonable at leading order in $\frac{1}{N}$) that one need only consider holographic purifications, this result along with the discussion above is enough to guarantee the subregion complexity = purification complexity, provided one accepts complexity = volume or complexity = spacetime volume for geometries dual to pure states.

### A. Maximal volumes of subregion wedges

Here we will give a proof of superadditivity of holographic subregion complexity in CV. Let $w_X$ denote the entanglement wedge [24] of a given boundary subregion $X$ of a boundary time slice, and let $\Sigma_X$ denote the maximal volume slice of $w_X$. By the maximal slice of $w_X$, we follow the proposal in [13], according to which we maximize over the volumes of slices anchored at both $X$ and the HRT surface of this boundary subregion. Given nonoverlapping subregions $A$ and $B$ on a Cauchy slice of the boundary, we have the inequality

$$\text{Vol}(\Sigma_{AB}) \geq \text{Vol}(\Sigma_A) + \text{Vol}(\Sigma_B). \qquad (16)$$

That this inequality is obeyed can be easily seen as follows: First, let us note that because $A$ and $B$ are nonoverlapping, it is also true that $w_A$ does not overlap $w_B$ (see [9]). Now either $\Sigma_A$ and $\Sigma_B$ meet to form a spatial slice of $w_{AB}$, or there is a gap between them. An example of the first case is a two-sided black hole, with $A$ and $B$ taken to be the entire left and right boundary respectively (see Fig. 6 for an illustration). An example of the second case is two entanglement wedges in pure AdS, the union of which is not the whole bulk (see Fig. 7 for an illustration). In the case without a gap the inequality is immediate: $\Sigma_{AB}$ is the *maximal* spatial slice over $w_{AB}$, so the slice formed by the union of $\Sigma_A$ and $\Sigma_B$ (whose volume is the sum of the individual volumes) must have a volume which does not exceed that of $\Sigma_{AB}$. In the case where there is a gap between $\Sigma_A$ and $\Sigma_B$ in $w_{AB}$, we may bridge this gap with a slice $\Sigma_{\text{bridge}}$ of $w_{AB} - w_A - w_B$ which meets $\Sigma_A$ and $\Sigma_B$ at their boundaries. We will moreover require that $\Sigma_{\text{bridge}}$ contains the HRT surface for $AB$. Then it is clear that the union of $\Sigma_A$, $\Sigma_B$, and $\Sigma_{\text{bridge}}$ forms a Cauchy surface for $w_{AB}$. But once again, since $\Sigma_{AB}$ is maximal, we have that

$$\text{Vol}(\Sigma_{AB}) \geq \text{Vol}(\Sigma_A) + \text{Vol}(\Sigma_B) + \text{Vol}(\Sigma_{\text{bridge}})$$
$$\geq \text{Vol}(\Sigma_A) + \text{Vol}(\Sigma_B), \quad (17)$$

where the second inequality holds because volume is a non-negative quantity. This establishes the superadditivity property for maximal volumes of subregion wedges.

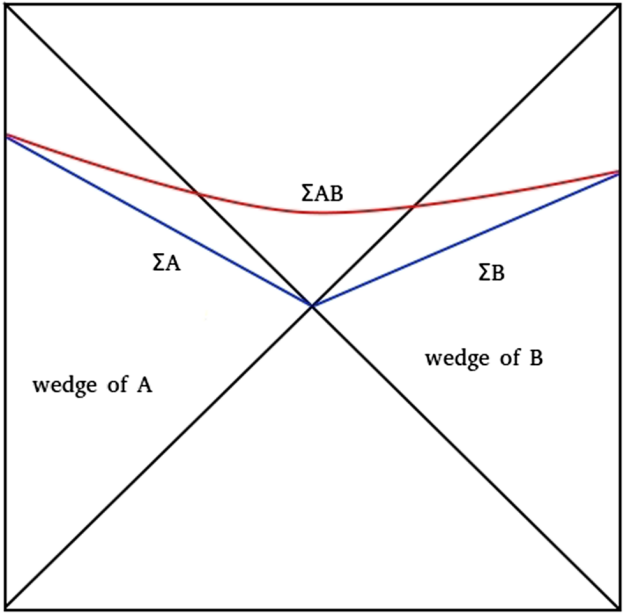

FIG. 6. A slice through entanglement wedges who share an HRT surface, as in the case of a two-sided BH, or the wedges of two halves of the boundary of pure AdS.

Notice that it is important to require that the bridge contains the HRT surface for $AB$. This is because the volume that computes the complexity is only maximal among Cauchy surfaces of the entanglement wedge, i.e., among surfaces anchored on this latter HRT surface (and also on $AB$). That $\Sigma_{\text{bridge}}$ can be chosen in this way follows from the results in [25], according to which it is always possible to choose a spatial slice in the bulk containing simultaneous the HRT surfaces for $A$, for $B$ and for $AB$ (assuming Einstein gravity together with the null energy condition).

### B. CV 2.0

Superadditivity also holds for the spacetime volumes of Wheeler-DeWitt patches, which has also been proposed as the dual quantity to state complexity in [7].

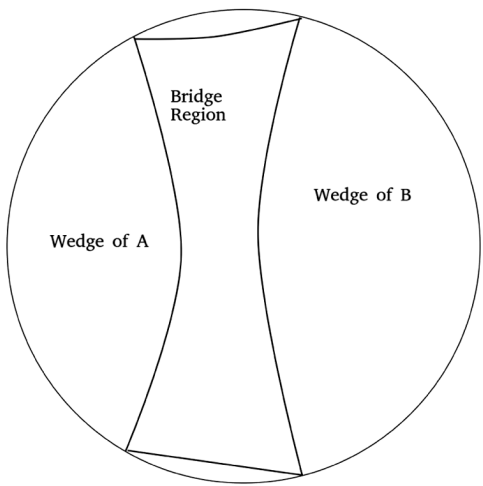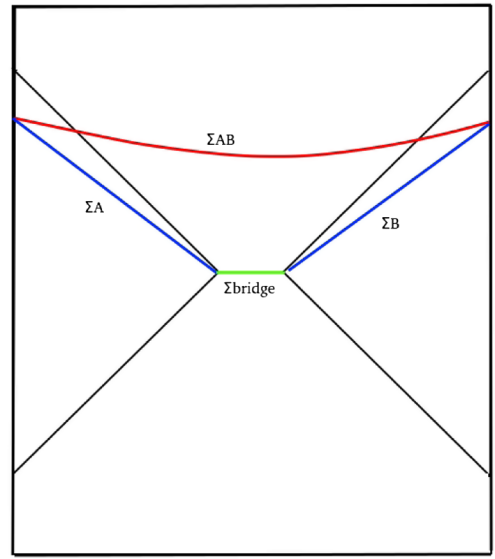

FIG. 7. On the left is a spatial slice showing the wedge of a region $A$, the wedge of a region $B$, and a "bridge region." Here there is clearly a gap between a spatial slice on the $A$ wedge and a spatial slice in the $B$ wedge, as shown in the cross section on the right. We can however always bridge this gap by an arbitrary spatial slice of the bridge region which meets the slices associated to A and B respectively at their boundaries.

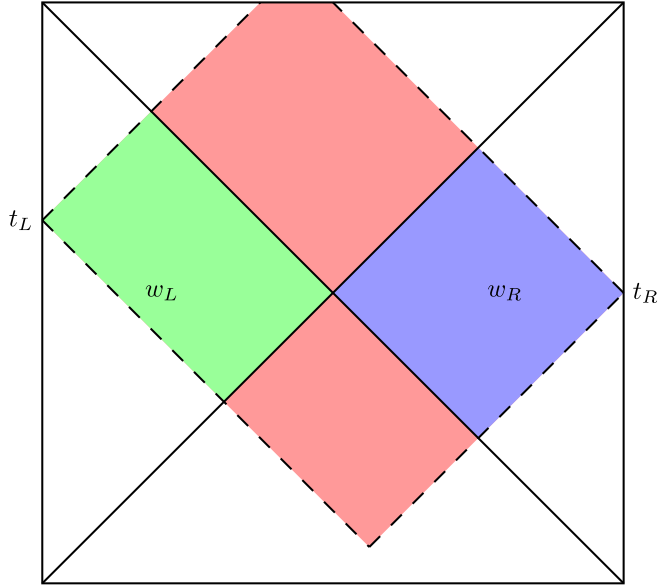

FIG. 8. The WDW patch for a two-sided black hole, outlined with dashed lines. If we consider subregion $A$ to be the left boundary and subregion $B$ to be the right boundary, then the entire outside of the horizon on the left side, $w_L$, is the entanglement wedge of $A$, and likewise for the right side and $B$. The entanglement wedge of $AB$ is the whole spacetime.

The superadditivity of WDW patch volumes follows trivially from the fact that given boundary regions $A$ and $B$, the entanglement wedges $w_A$ and $w_B$ of these regions are both subsets of the entanglement wedge $w_{AB}$ of the combined system. Hence, it is also true that the intersections of the WDW patch with $w_A$ and $w_B$ respectively are contained in the intersection with $w_{AB}$. Because spacetime volume is additive, this means the subregion complexity on $A$, $B$, and $AB$ respectively, which are given by the volumes of the intersections of the WDW patch with $w_A$, $w_B$, and $w_{AB}$, obey inequality (15). This is illustrated for a two-sided black hole in Fig. 8, where region $A$ is taken to be the whole left boundary and region $B$ is likewise taken to be the right boundary. Then the subregion complexity of $A$ is given by the spacetime volume of the region shaded in green, that of $B$ by the spacetime volume of the region shaded in blue, and the complexity on $AB$ is given by the spacetime volume of the whole WDW. The difference $\mathcal{C}_{AB} - \mathcal{C}_A - \mathcal{C}_B$ is thus given by the spacetime volume of the region shaded in red, which is clearly a positive definite quantity.

### C. Additivity in CA?

Since we have not proved superadditivity in either CA or CA-2, it is natural to wonder under what conditions, if any, such a property holds for these conjectures. We do not pursue this question in depth here, but it is quick work to see that superadditivity is far from generic in either proposal. The spacetime regions considered under the complexity = spacetime volume proposal are identical to those considered in the action proposals, namely, the intersection of a region's entanglement wedge and the WDW patch of the boundary slice. The same nesting properties then hold here as in the previous section. However, the fact that volume and spacetime volume are positive-definite quantities led to the superadditivity conditions above. On the other hand, the bulk integral of the action is often negative. For instance the Einstein Hilbert term in pure AdS spacetimes gives a negative integrand ($R - 2\Lambda = \frac{-2(d-1)}{L^2}$ in $d$ spacetime dimensions). There are also contributions from subregion boundaries and boundary intersections which may be of either sign. Some of these contributions have ambiguities which require stating a convention before the overall sign could be determined [13,14,26]. In the next section, we will look at specific instances where both subadditivity and superadditivity occur for the action.

## V. ADDITIONAL PURIFICATIONS

In Sec. III B we argued that assuming the complexity = volume or complexity = spacetime volume prescription for pure states, then the corresponding subregion prescription computes the purification complexity of the boundary mixed state, specifically minimizing complexity overall holographic purifications. The consistency conditions mentioned in II D follow automatically for these proposals. Similar demonstrations for subregion action prescription are not forthcoming, so in this section, we put the action subregion prescription to the test holographically by considering what is perhaps the most canonical holographic mixed state, the thermal state dual to one size of a static eternal black hole.

### A. Multisided black holes

In $2+1$ spacetime dimensions, the entanglement wedge corresponding to the thermal mixed state under consideration amounts to one exterior region of a BTZ black hole. The most obvious purifying geometry is the two-sided eternal black hole solution, dual to the thermofield double state, and we will scrutinize this case more carefully in the next subsection. However, a much larger class of holographic purifying geometries exists, which we consider all together here: $AdS_3$ black holes with $n$ sides and genus $g$. These geometries are an extension of the familiar BTZ black hole [27], obtained as a quotient of pure $AdS_3$ by a discrete group of isometries, and they are studied in [28–30]. The states dual to these geometries were discussed in [31], and the entanglement structure was studied in [32]. The minimum complexity over all sides $n$ and all genus $g$ provides an upper bound on the purification complexity of the state on a single boundary.

The complexity of this family of black holes was computed for the $t_1 = t_2 = \ldots = t_n = 0$ spatial slice in [15] according to both the complexity = volume and

complexity = action conjectures. The authors of that paper computed a quantity $\Delta C$, which they defined as the difference in complexity between the solution of interest, and $n$ copies of the one-sided $M = 0$ BTZ black hole. The results they found for an $n$ sided black hole with a wormhole of genus $g$ are

$$\Delta C_A = \frac{1}{6} c\chi, \tag{18}$$

$$\Delta C_V = -\frac{4}{3}\pi c\chi, \tag{19}$$

where $c = \frac{3L}{2G}$ is the central charge of the boundary CFT and $\chi = 2 - 2g - n$ is the Euler characteristic of the $t_1 = t_2 = \ldots = t_n = 0$ spatial slice. One immediate consequence of this result noted in [15], is that at any fixed $n$, $\Delta C_A$ decreases with increasing $g$. This already casts doubt on the idea that purification complexity should be identified with subregion complexity in CA, as for any value of the subregion complexity (which certainly does not depend on the genus, a property of the purification), we may find a genus such that the corresponding 2-sided purification has a lower complexity. This would seem to be a problem for complexity = action generally, in so far as it implies the purification complexity of our state in CA is $-\infty$. The authors of [15] suggest that perhaps this merely indicates an upper bound on the allowed genus of such black holes. In order to remain consistent with purification complexity = subregion complexity, however, one would need an even stricter bound than that implied by the positivity of complexity, or else our upper bound on the purification complexity will fall below the subregion complexity, the presumed true value.

By contrast, we see that for complexity = volume, increasing either the genus or the number of sides will only increase the complexity of the corresponding purification, and so our upper bound is provided by the ordinary BTZ case where $n = 2$ and $g = 0$. We see that in that case that $\chi$ vanishes, and so the total complexity is twice that of the $M = 0$ black hole (this calculation, of course, did not include any boundary term as suggested in Sec. III C). On the other hand, due to the mass independence of the subregion complexity of one side of BTZ, the subregion complexity is identical to that of one side of the $M = 0$ black hole, and so for CV we have that our upper bound on purification complexity is still larger than the subregion complexity, consistent with the conjecture that purification complexity = subregion complexity in CV.

This comes as no surprise, as this result was guaranteed by superadditivity. Given any subregion $A$ of any asymptotically AdS geometry, we may partition the total boundary into $A$ and its complement. Then superadditivity of CV, along with the positivity of volume, guarantees us that the holographic complexity of the total state (as computed in CV) is greater than the subregion complexity of $A$ (or equal to, in the limiting case where $A$ is the full boundary). Thus, given a state $\rho$ on a subregion $A$ of a CFT, and all classical geometries dual to purifications of $\rho$, i.e., all geodesic completions of the entanglement wedge $W$ of $A$ which preserve $W$ as the entanglement wedge of $A$, we will always find that according to CV, our holographic estimation of the purification complexity of $\rho$ is no greater than the subregion complexity of $A$.

Because the spacetime volume is also strictly positive, and because complexity according to CV2.0 is also superadditive, the same logic applies. Whether it applies to CA-2 in cases where it disagrees with CV2.0 is left to future work.

### B. Two-sided BTZ black hole: Detailed treatment

Though the action results for the genus $g$ black hole solutions may cast doubt on the idea that subregion action is dual to purification complexity, they could alternatively be teaching us nontrivial information about limits on the genus which can be described holographically at a certain cutoff, or these solutions might be disallowed for some other unknown reason. With these possibilities in mind, we examine in more depth the standard case of a two-sided (genus zero) BTZ black hole. The metric is given by

$$ds^2 = \frac{L^2}{z^2}(-f(z)dt^2 + f(z)^{-1}dz^2 + L^2 d\theta^2),$$

$$f(z) = 1 - \left(\frac{z}{z_H}\right)^2$$

with $z \to 0$ representing the AdS boundary and $z_H$ being the $z$ coordinate of the horizon. Subregion action calculations for this case were computed in [14,33], but we here report results for the total state complexity where the cutoff scale on left and right side are allowed to vary independently. We restrict our attention to the $t_L = t_R = 0$ slice.

One convention for action computations chooses null boundary generators to be affinely parametrized with arbitrary normalization constants. These free parameters may have interesting interpretations in terms of boundary theory parameters or particular notions of quantum state complexity ([13,34]). In the Appendix we report results under these conventions, but the action so defined is obviously not invariant under reparametrization of the boundary generators. For this reason the results reported here include the counterterm identified in Appendix B of [26], which does render the action reparametrization invariant. This counterterm eliminates dependence on the choices above but instead requires the introduction of an undetermined length scale on each null boundary. In principle these length scales can differ on ingoing/outgoings null boundaries, but only the product of these appears in action results. For simplicity we here set them equal and denote the single length scale $\tilde{L}$. See Appendix for full expressions in both conventions.

We utilize cutoff surfaces of constant $z < z_H$. On the right and left sides denote these $z_{R,\min} = \delta_R$ and $z_{L,\min} = \delta_L$, respectively. We report the action result for the $t_L = t_R = 0$ slice here, abbreviating $\bar{L} = \tilde{L}/L$ where $L = L_{\mathrm{AdS}}$, $\bar{\delta}_L = \delta_L/z_H$, and $\bar{\delta}_R = \delta_R/z_H$:

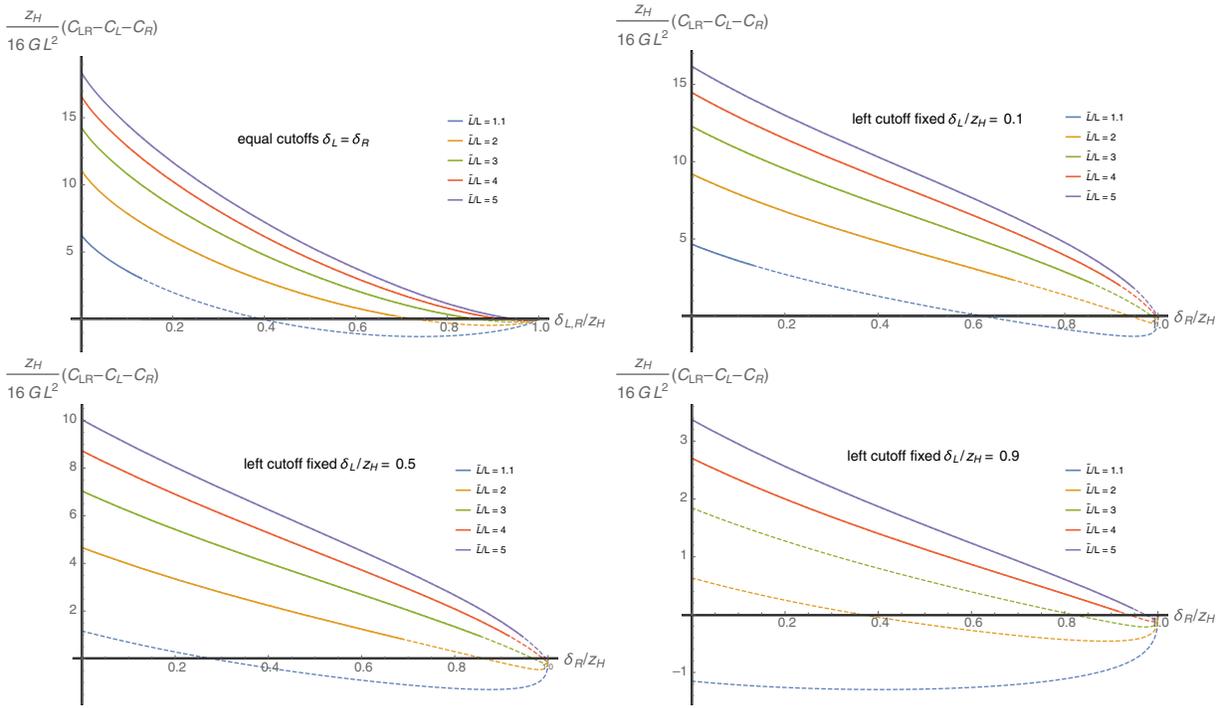

FIG. 9. Plots show the difference between the total state complexity $C_T = C_{LR}$ and the sum of left and right subregion complexities $C_L + C_R$ for the BTZ black hole at $t_L = t_R = 0$, computed using the complexity = action prescription. Positive values indicate that the left and right factors behave superadditively, and negative indicates subadditivity. The cutoffs are constant $z$ surfaces $z_R = \delta_R$, $z_L = \delta_L$ and various choices of $\tilde{L}/L_{\text{AdS}}$ are displayed. Positivity in the strict UV limit ($\delta_L \to 0$ and $\delta_R \to 0$) requires that the length scale $\tilde{L}$ be chosen greater than $L_{\text{AdS}}$, so only such lines are displayed. The dashed portions of each line are excluded if we demand positivity of the total state complexity as well as both subregion complexities at the corresponding cutoffs. We find that in the remaining parameter space the system always behaves superadditively: $C_T > C_L + C_R$.

$$\frac{\mathcal{A}^T(\bar{\delta}_L, \bar{\delta}_R)}{16\pi G} = \frac{2L^2}{z_H}\left(-\frac{2(s_L s_R - 1)\log\left(\frac{4\bar{L}^2 s_L s_R}{(s_L s_R - 1)^2}\right)}{s_L s_R + 1}\right.$$
$$\left. + \frac{\log((1-\bar{\delta}_L^2)\bar{L}^2)}{\bar{\delta}_L} + \frac{\log((1-\bar{\delta}_R^2)\bar{L}^2)}{\bar{\delta}_R}\right) \quad (20)$$

where $s_L = \sqrt{\frac{1+\bar{\delta}_L}{1-\bar{\delta}_L}}$ and $s_R = \sqrt{\frac{1+\bar{\delta}_R}{1-\bar{\delta}_R}}$. If the cutoffs are taken to be symmetric, this simplifies to

$$\frac{\mathcal{A}^T(\bar{\delta}, \bar{\delta})}{16\pi G} = \frac{4L^2}{z_H}\left(-\bar{\delta}\log\left(\bar{L}^2\left(\frac{1}{\bar{\delta}^2} - 1\right)\right)\right.$$
$$\left. + \frac{1}{\bar{\delta}}\log(\bar{L}^2(1-\bar{\delta}^2))\right). \quad (21)$$

The subregion action associated with either left or right side (restricted to the entanglement wedge[11]) is given by

$$\frac{\mathcal{A}^{L,R}(\bar{\delta})}{16\pi G} = \frac{2L^2}{z_H}\left(-\log\left(4\bar{L}^2\frac{(1-\bar{\delta})}{(1+\bar{\delta})}\right) + \frac{1}{\bar{\delta}}\log(\bar{L}^2(1-\bar{\delta}^2))\right). \quad (22)$$

Comparing these results to expectations is more subtle because of the existence of the undetermined length scale $\tilde{L}$, which lacks a clear interpretation in relation to complexity in the boundary theory. Demanding that the complexity is a positive quantity in the strict UV limit ($\delta_L \to 0$, $\delta_R \to 0$) simply requires that we choose $\tilde{L} > L_{\text{AdS}} \Rightarrow \bar{L} > 1$. If the interpretations of Sec. III B are correct, however, and letting the cutoffs on both sides approach the horizon ($\delta \to z_H$ or $\bar{\delta} \to 1$) amounts to coarse-graining the state on both boundaries, then positive complexity should be imposed over the whole range of allowed cutoffs. We find that $\bar{L}$ greater than but of order 1 easily results in negative complexity as the cutoffs are pushed inward (away from $z = 0$). For $\bar{L} \gg 1$, both cutoffs must approach the horizon itself before there is a problem with negative complexity. For any choice of cutoffs, there is some $\bar{L}$ sufficiently large to avoid negative complexity. This is especially true if at least one cutoff remains in the deep UV (e.g., $\bar{\delta}_R \ll 1$).

If we ignore positivity requirements and allow any combination of fixed-$z$ cutoffs, then we can find regions of subadditivity ($C_T < C_L + C_R$), as well as regions where the total state complexity is lower than one or the other of the subregion complexities ($C_T < C_L$ or $C_T < C_R$), which would violate the purification complexity interpretation of the corresponding subregion action. Neither of these properties occurs, however, if we restrict to regimes where the total state complexity as well as both subregion complexities are positive. See Fig. 9 for illustration of how requiring positivity excludes cases of subadditivity, for example.

---

[11]Joint contributions in the subregion action diverge if naively evaluated on the black hole horizon. We followed [14] by computing the action on an exterior region bounded by null surfaces just outside the future and past horizons and taking the limit that they coincide with the future and past horizons.

To summarize the action results for this case, if we treat positivity as a strict condition for all complexities, then the choice of length scale $\tilde{L}/L_{\text{AdS}}$ can impose nontrivial restrictions on the allowable cutoff surfaces (especially for $\tilde{L}/L_{\text{AdS}}$ greater than but of order one). Over any combination of $\tilde{L}/L_{\text{AdS}}, \delta_L, \delta_R$ which gives positive complexities, we find that the total state complexity is always larger than either subsystem complexity, avoiding contradiction with the purification complexity = subregion action conjecture as could occur for the genus $g$ solutions of the previous section. In the parameter space consistent with positivity, the action behaves superadditively. As reviewed in Sec. II C, this differs from the subadditivity expectation coming from the basis and spectrum decomposition utilized in [14]. The discussion of that section cast some doubt on the subadditivity expectation, but we nevertheless found it plausible. The mixed results of this section do not give any strict violation of the purification complexity interpretation of subregion action, but without a reason to expect superadditivity as the generic behavior for these states it remains a puzzle facing that interpretation. The significance of $\tilde{L}$ is an important piece in that puzzle.

If we consider instead the subregion volume prescription for the same spacetime decomposition, including the boundary term on the HRT surface suggested in Sec. III C), then at least our most immediate expectations follow naturally. All complexities are manifestly positive, and subadditivity is always obeyed on the $t_L = t_R = 0$ slice. The sum of the subregion complexities "double counts" the boundary term, which the total state slice effectively counts only once. Because the boundary term is proportional to the area of the HRT surface (and therefore the entanglement entropy) and it does not vanish under any course graining, we loosely interpret it as the "basis complexity" portion of the subregion complexity.

The volume prescription also incorporates the case we puzzled over after Eq. (11), where the total state happens to be exactly the minimal complexity purification of both subsystems, and we saturate the weak superadditivity limit $C_L + C_R = 2C_T$. This would not ordinarily occur for the thermofield double state unless it happened to be the minimally complex purification for both left and right factors. This can be realized as a limiting case, saturated by taking both cutoffs down to their limiting values around the HRT surface (equivalently, sending the volume portion to zero and leaving only the HRT boundary term, refer to Fig. 4). This could perhaps be thought of as effectively course grains the basis complexity on both sides to zero and leaves only the spectrum part, necessary to maintain the entanglement structure and give a pure state.

### C. Subregions of pure AdS$_3$

Finally, we will consider the reduced state on interval subregion of pure AdS. Actually, the subregion complexity of general subregions of AdS$_3$ was computed according to CV in [35]. It remains, however, to repeat that computation for CA, and to compute the purification complexity of such subregions, by considering all (or at least a large family of) geodesic completions of the entanglement wedge. The state on any particular interval subregion in pure AdS$_3$ is related to the state on other interval subregions, whether of the global boundary or the Poincaré boundary, by boundary conformal transformations, and the entanglement wedge to which it is dual can be related to other entanglement wedges by bulk isometries. The entanglement wedge may also be isometrically mapped to an entanglement wedge of a subregion of the boundary of the BTZ black hole, or more generally to a subregion of a single side of the $n$ sided genus $g$ black hole. These transformations, however, will not, in general, preserve the cutoff, and so we must keep track of how the cutoff transforms under such transformations.

To begin with, we will consider an interval of length $2R$ along with a cutoff defined by constant Poincaré bulk coordinate $z$, namely, we will put a hard cutoff at $z = \delta$. The metric in this scenario is the familiar

$$ds^2 = \frac{L^2}{z^2}(-dt^2 + dx^2 + dz^2). \qquad (23)$$

One purifying geometry is of course the entire Poincaré patch, the complexity of which can be computed to be (according to CA)

$$C = \frac{V_x L}{4\pi^2 \hbar G \delta}\left[\log\left(\frac{\tilde{L}}{\delta}\right) - 3\right] \qquad (24)$$

where $L$ is the AdS length scale and $\tilde{L}$ is an arbitrary length scale. Periodically identified Poincaré space, where we identify $x$ with $x + a$, also provides a purifying geometry for our interval, provided the periodic identification does not change the RT surface. This requires that the length of our interval must not be greater than half the length of the periodically identified boundary. One may see this as follows: Consider a region of length $a$ on the Poincaré patch periodically identified as $x \sim x + R$. Clearly, for $\frac{a}{R}$ sufficiently small, the entanglement wedge of the region is the same as in the full Poincaré patch. How small though does $\frac{a}{R}$ need to be? In the periodically identified geometry, our interval from $x = 0$ to $x = a$ is homologous to the interval from $x = a$ to $x = R$. As a consequence, there are two surfaces competing to be the global minimum, one which lifts in the full Poincaré patch to the usual RT surface for the interval of interest, and one which lifts instead to the RT surface of the interval from $x = a$ to $x = R$. Clearly, the "usual" RT surface wins out whenever $a < R - a$, i.e., when $\frac{a}{R} < \frac{1}{2}$. Because it is clear that the complexity of such a purification grows monotonically with $R$, the lowest complexity in this family will be given by $R = 2a$, right at the point of transition between the two minima. The complexity of this periodically identified geometry will be the same as for the nonidentified case, but with the infinite transverse volume simply replaced by $2R$, yielding

$$C^P(R) \leq \frac{LR}{2\pi^2 \hbar G \delta} \left[ \log\left(\frac{\tilde{L}}{\delta}\right) - 3 \right] \quad (25)$$

Notice now that if we scale all the coordinates, as well as the size of our region and our cutoff, by a uniform constant $\alpha$, that is an isometry, such that the image of our original cutoff entanglement wedge is the new cutoff entanglement wedge (with the new cutoff). If one does this while holding $\tilde{L}$ fixed, one may, in fact, cause the complexity of the regulated complexity to become negative. This may be avoided, however, if we scale $\tilde{L}$ along with the coordinates too, and the purification complexity bound found above is fixed under such rescalings.

We may also embed such regions as subregions of a single side of a BTZ black hole, or even to an entire side of a planar BTZ black hole (although the corresponding cutoff will not appear natural). To see this, first, consider that the entanglement wedge of an interval on the Poincaré patch may be mapped by a simple change of coordinates to one side of AdS-Rindler. By applying the map corresponding to our region of interest, we can get our entanglement wedge as one side of AdS-Rindler, and by considering a larger interval containing ours, we may get our interval as a subregion of the boundary of one side of AdS-Rindler. Notice that the AdS-Rindler metric may be written as

$$ds^2 = \frac{L^2}{z^2} \left[ -f(z) dz^2 + \frac{dz^2}{f(z)} + L^2 d\chi^2 \right] \quad (26)$$

where $f(z) = 1 - \frac{z^2}{L^2}$. This is exactly the same metric as the planar BTZ black hole with $z_h = L$. Replacing $f(z)$ by $f(z) = 1 - \frac{z^2}{z_h^2}$ then, we have AdS-Rindler when $z_h = L$, but the metric is invariant under scaling all the coordinates along with $z_h$ (but not the AdS length scale $L$) by any constant $\alpha$. The rescaling gives us an isometry between AdS-Rindler and a BTZ black hole of any temperature. Composing the two isometries we have found, we may then embed our regulated entanglement wedge either as a full side of a BTZ black hole or as the entanglement wedge of a subregion of a single boundary. We may also see that the entanglement wedge of our interval may be embedded into BTZ by recalling that BTZ black holes (as a specific example of the $n$-sided genus $g$ construction discussed above) may be obtained by a quotient of pure global AdS$_3$. Clearly then, any interval subregion of the boundary of AdS$_3$ may be mapped by a conformal transformation to an interval lying entirely inside a fundamental domain of our quotient space. Because such subregion lies entirely in a fundamental domain, that interval may be thought of as a subregion on the boundary of the quotient space. Either way, we may minimize the resulting embedding over a number of parameters, such as the size of the BTZ subregion or the BTZ temperature, but first, let us come back to the $n$ sided genus $g$ solutions discussed above. Clearly, for any embedding into a single side of a BTZ geometry, these also form a family of geodesic completions of the entanglement wedge of our interval. However, the conclusion from considering this family is no different from those we already saw for the BTZ thermal state. For CV and CV2.0, the complexity increases both with the number of sides $n$ and the genus $g$. For CA, the complexity increases in $n$ (at least for a sufficiently small choice of the cutoff), but decreases without bound with increasing $g$. This leads to the same problems as before and does not tell us anything particularly new.

## VI. CONCLUSION

In this paper, we have estimated the purification complexity of the reduced state on a boundary subregion holographically, by considering different geodesic completions of the dual entanglement wedge, and by considering different cut-offs, corresponding to different course-grainings of the purifying state. We have shown that according to CV and CV2.0, the optimal holographic purification is given by a geodesic completion of the entanglement wedge with a cutoff which skirts the HRT surface of the compliment of our region. As such, the minimum complexity among holographic purifications reproduces the usual subregion complexity (possibly up to an extra boundary term, as discussed in Sec. III C). As such, we may conclude that at least to leading order in the $\frac{1}{N}$ expansion, subregion complexity = purification complexity if either CV or CV2.0 is correct for geometries dual to pure states.

For CA, we still find that the subregion complexity corresponds to the complexity of a particular holographic purification, though we have no guarantee that there is not some other purification of smaller complexity. In fact, we have several explicit examples in which what naively appears to be a holographic purification is found to have a smaller complexity than the subregion complexity in question, and even that the complexity can become negative. From this we must choose from the following conclusions, namely that these combinations of geometry and cutoff are not genuinely holographic (i.e., they are not dual to some state in a boundary CFT), that the complexity = action conjecture is simply wrong, or that the complexity = action is a special case of a more general conjecture, so that in these cases CA does not apply and we must use the generalization. As an example of the last situation, one might imagine adding a term which depends explicitly on the Euler characteristic in order to "fix" CA in case of arbitrary behind the horizon genus.

These results are in some tension with those found in [14]. In the case of CV, this tension is not so strong. While it is true that CV is superadditive (as the authors of that paper had pointed out), we argue that in most circumstances this is to be expected due to the existence of operations which can raise the complexity of a pure state while going unnoticed by the density matrices arising from a decomposition into subsystems. In the particular case that preparing the total state does not require any such operations (as may be the case for the $t_L = 0$, $t_R = 0$ thermofield double state decomposed on right and left factors), then subadditivity can be realized modifying the prescription for subregion complexity in CV by adding the boundary term discussed in Sec. III C. Once this modification is made, the arguments made in [14] that subregion complexity in CA matches purification complexity applies equally well to CV. For CA, the tension is perhaps stronger, though it mostly arises from cases not considered by those authors.

## ACKNOWLEDGMENTS

We would like to thank Cesar Agón, Adam Brown, Shira Chapman, Matt Headrick, Ted Jacobson, Phuc Nguyen, Eliezer Rabinovici, Henry Stoltenberg, and Brian Swingle for useful discussions and feedback. In particular, part of this paper grew out of earlier work with Ted Jacobson and Phuc Nguyen. J. C. and S. E. would like to thank IAS for their hospitality during PITP 2018 and for thought-provoking talks and conversation. This work was supported by NSF Grants No. PHY-1620610 and No. PHY-1820712.

## APPENDIX: CALCULATION OF TWO-SIDED BTZ ACTION

We here calculate the on shell action for the BTZ black hole with metric described by

$$ds^2 = \frac{L^2}{z^2}(-f(z)dt^2 + f(z)^{-1}dz^2 + L^2 dx^2)$$
$$f(z) = 1 - \left(\frac{z}{z_H}\right)^2.$$

We will allow the cutoffs on the left and right side to vary independently, although we still choose constant $z$ surfaces, and we consider only the $t_L = t_R = 0$ slice. We denote these cutoffs as $z_L = \delta_L$, $z_R = \delta_R$, although they are not necessarily small.

We will consider the total WDW patch action ($\mathcal{A}^T$), as well as that of the right subsystem ($\mathcal{A}^R$). The latter is obtained by restricting to the entanglement wedge of the subsystem, which is simply the right exterior of the black hole. The left subsystem action ($\mathcal{A}^L$) is the same as $\mathcal{A}_R$ with $\delta_R \to \delta_L$. Time reflection symmetry and additivity of the bulk action also allow for simple determination of the action on the behind-the-horizon regions to the future ($\mathcal{A}_F$) and past ($\mathcal{A}_P$) through $\mathcal{A}_F = \mathcal{A}_P = \frac{1}{2}(\mathcal{A}_T - \mathcal{A}_R - \mathcal{A}_L)$. These two bulk subregions are not expected to correspond to boundary state complexities in their own right so we do not compute them here.

The formalism for computing the action on spacetime regions with null boundaries has only recently received attention [26,36–39]. We will first use the prescription summarized in Appendix C of [26] with the generators of null boundaries affinely parametrized. This leaves a dependence on the overall normalization of these generators [see Eq. (A3)]. We will also compute the boundary counterterms described in Appendix B of [26], the inclusion of which makes the action result reparametrization invariant. In this Appendix we report results with and without these terms.

### 1. Setup

Both the total state WDW patch and the left/right subregions we consider are "diamond shaped" on a conformal diagram, so we adopt a common labeling scheme for the boundaries and joints, shown in Fig. 10. In each case, the action $\mathcal{A}$ consists of a bulk contribution on the relevant volume $\mathcal{V}$, four contributions on null boundaries $\mathcal{N}_i$ ($i = 1, 2, 3, 4$), and four co-dimension two "joint" contributions.[12] We will denote these joints as $\mathcal{J}_j$ ($j = 1, 2, 3, 4$). We will also use sometimes use $j_1$ and $j_2$ to label the left-going/right-going null boundaries whose intersection forms the $j$th joint. For example for $j = 1$ the joint is the intersection of $\mathcal{N}_1$ and $\mathcal{N}_4$, so $(j_1, j_2) = (1, 4)$.

Before adding the term for reparametrization invariance, the action on such a bulk regions is given by

$$16\pi G \mathcal{A} = \mathcal{S} = \int_{\mathcal{V}} (\mathcal{R} - 2\Lambda)\sqrt{-g}dV$$
$$- 2\sum_i \text{sign}(\mathcal{N}_i) \int_{\mathcal{N}_i} \kappa dS d\lambda_i$$
$$+ 2\sum_j \text{sign}(\mathcal{J}_j) \int_{\mathcal{J}_j} \ln|k_{j_1} \cdot k_{j_2}/2| dS.$$

Here $\lambda_i$ parametrizes the $i$th null boundary through $k_i^\alpha = dx^\alpha/d\lambda_i$ with $\kappa$ defined by $k_i^\beta \nabla_\beta k_i^\alpha = \kappa k_i^\alpha$. We first choose this parameter to be affine, and express results in terms of arbitrary normalization constants for the $k_i$: $k_i \cdot \hat{t} = c_i$ with $\hat{t}$ being the static time coordinate vector $\hat{t} = \partial_t$. This leaves only the bulk and joint contributions, which we will label $S_{\mathcal{V}}$ and $S_{\mathcal{J}}$,

---

[12]Note that a prescription using variable cutoff surfaces to define a family of purifications is more naturally suited to a joint (constant $t$) cutoff scheme as opposed to a timelike boundary cutoff, see Fig. 11.

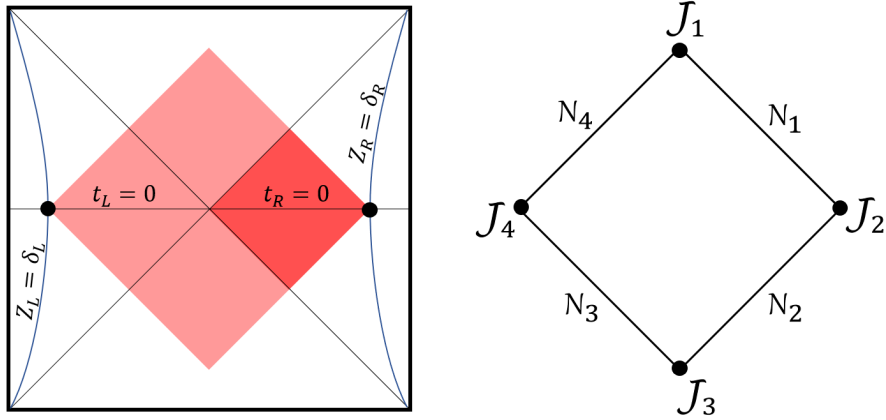

FIG. 10. A conformal diagram for the two-sided BTZ spacetime is shown in the left panel. The Wheeler-DeWitt patch corresponding to the total state is shaded in light red, while dark red indicates the subregion of interest for the right subregion computation. Both regions are "diamond shaped" so we adopt a shared labeling scheme for the joints and null boundaries, as illustrated in the right panel.

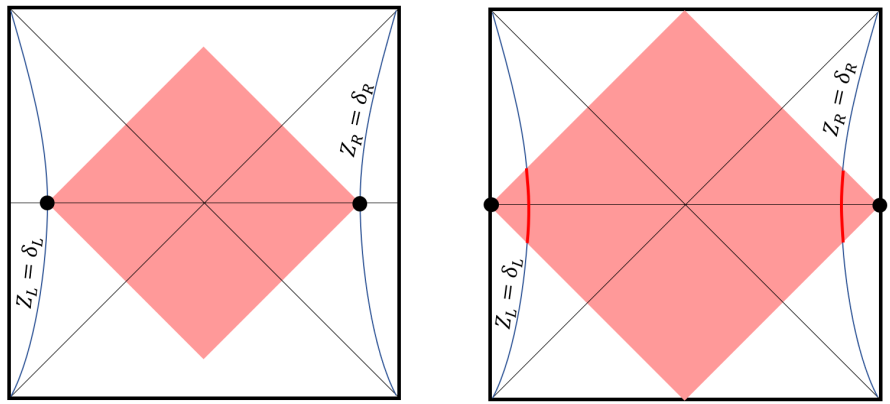

FIG. 11. The cutoff prescription utilized in this paper is that of the left diagram.

$$S_\mathcal{V} = \int_\mathcal{V} (\mathcal{R} - 2\Lambda)\sqrt{-g}\, dV,$$

$$S_\mathcal{J} = 2\sum_j \text{sign}(\mathcal{J}_j) \int_{\mathcal{J}_j} \ln|k_{j_1} \cdot k_{j_2}/2|\, dS.$$

After computing these contributions we will add counter-terms associated with the null boundaries which make the action reparametrization invariant. These terms ultimately render the choices of the previous paragraph irrelevant (the affine parameterization and normalization of $k_i^\alpha$), but they require the introduction of arbitrary length scale(s) which we will denote $\tilde{L}_i$.

For the $i$th null boundary, the counterterm $\Delta S_{\mathcal{N}_i}$ is given by

$$\Delta S_{\mathcal{N}_i} = 2\,\text{sign}(\mathcal{N}_i) \int_{\mathcal{N}_i} \Theta_i (\ln|\tilde{L}\Theta_i|) dS d\lambda_i$$

where $\Theta_i = \frac{1}{\sqrt{\gamma}} \frac{d\sqrt{\gamma}}{d\lambda_i}$, the expansion parameter associated with $\lambda_i$. Here $\sqrt{\gamma}$ is the volume element induced on the space transverse to fixed $\lambda_i$.

The normalization constants $c_i$, and the length scales $\tilde{L}_i$ should be the same on future-directed left-going and right-going null boundaries (this ensures, for instance, that the action goes to zero when the "diamond" is squashed along either null direction). At the end we will impose this but we will first denote them independently to illustrate more explicitly the replacement of $c_i$ dependence by $\tilde{L}_i$ dependence. The signs for both null boundary and joint contributions will be listed in their respective subsections below.

#### 2. Bulk contribution

The bulk contributions on each subregion simply integrate over the constant value of $R - 2\Lambda = -4/L^2$. The spacetime volume of each wedge (right/left/future/past) can be written $\mathcal{V} = \Delta x L^4 |\frac{1}{z_c} - \frac{1}{z_H}|$, where $z_c$ is the position of the joint furthest from the bifurcation surface in each quadrant. In the right and left quadrants, $z_c$ is simply the cutoff $z_c = \delta_{L,R}$. In the future and past quadrants we denote these joints as $z_F$ and $z_P$, though symmetry dictates that these are at the same coordinate $z$ value, determined by the two cutoffs as

$$z_F = z_P = z_H \frac{1 + \sqrt{\frac{(z_H+\delta_R)(z_H+\delta_L)}{(z_H-\delta_R)(z_H-\delta_L)}}}{-1 + \sqrt{\frac{(z_H+\delta_R)(z_H+\delta_L)}{(z_H-\delta_R)(z_H-\delta_L)}}}. \quad (A1)$$

The bulk contribution in each quadrant can therefore be written in the unified form

$$S_\mathcal{V}^{R,L,F,P} = -4\Delta x L^2 \left| \frac{1}{z_{R,L,F,P}} - \frac{1}{z_H} \right|. \quad (A2)$$

### 3. Joint contribution

Turning to the joint contributions, we utilize affinely parametrized future-directed boundary generators, normalized according to $k_i \cdot \hat{t} = -c_i$ where $\hat{t} = \partial_t$. For example in the right quadrant we choose future-directed ingoing and outgoing generators on $\mathcal{N}_1$ and $\mathcal{N}_2$ to be given by

$$(k_1)_\mu = c_1 \partial_\mu(-t + z^*(z)),$$
$$(k_2)_\mu = c_2 \partial_\mu(-t - z^*(z)), \quad (A3)$$

where $z^*(z)$ is a BTZ "tortoise" coordintate:

$$z^*(z) = \frac{z_H}{2} \log\left| \frac{z + z_H}{z - z_H} \right|. \quad (A4)$$

Analogous expressions apply for all $k_i$. For every joint we will consider, we have $\log|k_{j_1} \cdot k_{j_2}/2| = \log\left(\frac{c_{j_1} c_{j_2} z^2}{L^2 |f(z)|}\right)$ and $dS = L^2/z\, dx$, so the total joint contributions can be written

$$S_\mathcal{J} = 2\Delta x \sum_j \left( \text{sign}(\mathcal{J}_j) \frac{L^2}{z_j} \ln\left( \frac{c_{j_1} c_{j_2} z_j^2 z_H^2}{L^2 |z_j^2 - z_H^2|} \right) \right). \quad (A5)$$

The signs of the joint contributions given by (see [26], Appendix C) $\text{sign}(\mathcal{J}_1)=1$, $\text{sign}(\mathcal{J}_2)=-1$, $\text{sign}(\mathcal{J}_3)=1$, $\text{sign}(\mathcal{J}_4) = -1$.

### 4. Counterterm

Lastly, we compute the counterterms required for reparametrization invariance. This requires the introduction of a length scale which we will denote $\tilde{L}_i$ on the null boundary $\mathcal{N}_i$. Once again these length scales should match for left-going and right-going boundaries, but we first denote them independently to keep track of contributions and cancellations with joints. For each null surface we have a contribution depending on $\Theta_i$, the expansion parameter associated with $\lambda_i$:

$$\Delta S_{\mathcal{N}_i} = 2\,\text{sign}(\mathcal{N}_i) \int_{\mathcal{N}_i} \Theta_i (\ln |\Theta_i \tilde{L}|) dS d\lambda_i.$$

For each of the four null surfaces in question, $\sqrt{\gamma} = (L^2/z)$ and so $\Theta_i = -\frac{1}{z}\frac{dz}{d\lambda_i}$. For the generators parametrized as in Eq. (A3), we have $dz/d\lambda_i = \pm c_i z^2/L^2$. This allows us to write each contribution in the form

$$\Delta S_{\mathcal{N}_i} = -2\,\text{sign}(\mathcal{N}_i)\Delta x \int_{z_i(\lambda_{\min})}^{z_i(\lambda_{\max})} \frac{L^2}{z^2} \log\left( \frac{c_i \tilde{L}_i z}{L^2} \right) dz$$
$$= 2\,\text{sign}(\mathcal{N}_i)\Delta x \frac{L^2}{z} \left( 1 + \log\left( \frac{c_i \tilde{L}_i z}{L^2} \right) \right) \Bigg|_{z_i(\lambda_{\min})}^{z_i(\lambda_{\max})}. \quad (A6)$$

With $\lambda$ integrated to the future, parametrization invariance requires $\text{sign}(\mathcal{N}_i) = -1$ for null boundaries which are to the future of the bulk subregion, and $+1$ for past null boundaries.

### 5. $\mathcal{A}^T$: Full WDW patch action

We first compute the full action on the WDW patch. Summing the bulk contributions from each quadrant we have

$$S_\mathcal{V}^T = -4\Delta x L^2 \left( \frac{1}{\delta_L} + \frac{1}{\delta_R} - \frac{1}{z_F} - \frac{1}{z_P} \right)$$

where $z_F = z_P$ are the future-most and past-most joint location, given by Eq. (A1). The joint contributions amount to

$$S_\mathcal{J}^T = 2\Delta x L^2 \left( \frac{1}{z_F} \log\left( \frac{c_1 c_4 z_F^2 z_H^2}{L^2(z_F^2 - z_H^2)} \right) + \frac{1}{z_P} \log\left( \frac{c_2 c_3 z_P^2 z_H^2}{L^2(z_P^2 - z_H^2)} \right) - \frac{1}{\delta_R} \log\left( \frac{c_1 c_2 \delta_R^2 z_H^2}{L^2(z_H^2 - \delta_R^2)} \right) - \frac{1}{\delta_L} \log\left( \frac{c_3 c_4 \delta_L^2 z_H^2}{L^2(z_H^2 - \delta_L^2)} \right) \right)$$

and the counterterms together give

$$\Delta S_\mathcal{N}^T = 4\Delta x L^2 \left( \frac{1}{\delta_L} + \frac{1}{\delta_R} - \frac{1}{z_F} - \frac{1}{z_P} \right) + 2\Delta x L^2 \left( \frac{1}{\delta_R} \log\left( \frac{c_1 c_2 \tilde{L}_1 \tilde{L}_2 \delta_R^2}{L^4} \right) + \frac{1}{\delta_L} \log\left( \frac{c_3 c_4 \tilde{L}_3 \tilde{L}_4 \delta_L^2}{L^4} \right) \right.$$
$$\left. - \frac{1}{z_F} \log\left( \frac{c_1 c_4 \tilde{L}_1 \tilde{L}_4 z_F^2}{L^4} \right) - \frac{1}{z_P} \log\left( \frac{c_2 c_3 \tilde{L}_2 \tilde{L}_3 z_P^2}{L^4} \right) \right).$$

The cancellation of $c_i$ constants from the joint terms after the inclusion of counterterms is evident from the above expressions. In the combined answers we impose that $c_1 = c_3 = c$, $c_2 = c_4 = \bar{c}$, $\tilde{L}_3 = \tilde{L}_1$, $\tilde{L}_4 = \tilde{L}_2$, as well as $z_P = z_F$. The total action, reported without ($\mathcal{A}^T$) and with ($\tilde{\mathcal{A}}^T$) the counterterms is therefore as follows:

$$\mathcal{A}^T = \frac{\Delta x L^2}{8\pi G}\left[\frac{2}{z_F}\left(2 + \log\left(\frac{c\bar{c}\, z_F^2 z_H^2}{L^2(z_F^2 - z_H^2)}\right)\right) - \frac{1}{\delta_R}\left(2 + \log\left(\frac{c\bar{c}\,\delta_R^2 z_H^2}{L^2(z_H^2 - \delta_R^2)}\right)\right) - \frac{1}{\delta_L}\left(2 + \log\left(\frac{c\bar{c}\,\delta_L^2 z_H^2}{L^2(z_H^2 - \delta_L^2)}\right)\right)\right],$$

$$\tilde{\mathcal{A}}^T = \frac{\Delta x L^2}{8\pi G}\left[\frac{2}{z_F}\log\left(\frac{L^2 z_H^2}{\tilde{L}_1\tilde{L}_2(z_F^2 - z_H^2)}\right) - \frac{1}{\delta_R}\log\left(\frac{L^2 z_H^2}{\tilde{L}_1\tilde{L}_2(z_H^2 - \delta_R^2)}\right) - \frac{1}{\delta_L}\log\left(\frac{L^2 z_H^2}{\tilde{L}_1\tilde{L}_2(z_H^2 - \delta_L^2)}\right)\right]. \tag{A7}$$

### 6. $\mathcal{A}^R$, $\mathcal{A}^L$: Subregion action

We now compute the right subregion action, defined by restricting the Wheeler-DeWitt patch to the entanglement wedge of the boundary subregion, which for the right boundary is simply the full right-side exterior of the black hole. In most respects the right subregion computation follows the same procedure as the previous section. The bulk contribution is given by Eq. (A2) as

$$S_V^R = -4L^2\Delta x\left(\frac{1}{\delta_R} - \frac{1}{z_H}\right).$$

However, evaluating joints directly on the horizon would naively give divergent results. We follow [14], by first setting null surfaces just outside the past and future horizons and taking the limit that they coincide with the horizons. More precisely, we define null boundaries $\mathcal{N}_3$ and $\mathcal{N}_4$ to pass through a joint just off the bifurcation surface: $t = 0$, $z = zH - \epsilon$ in the right quadrant (a different $t$ would not affect the result). We then let $\epsilon \to 0$, which ultimately leaves a finite result for the total joint contribution. Because of the simplicity of the $z^*(z)$ function in Eq. (A4), this procedure can be carried out exactly for any $\delta_R$.

Before setting the $c_i$ constants to $c$ and $\bar{c}$, the contribution from joints $\mathcal{J}_1$, $\mathcal{J}_3$, and $\mathcal{J}_4$ is

$$S_{\mathcal{J}_1}^R + S_{\mathcal{J}_3}^R + S_{\mathcal{J}_4}^R = \frac{2L^2\Delta x}{z_H}\left(\log\left(\frac{c_1 c_2 z_H^2(z_H + \delta_R)}{4L^2(z_H - \delta_R)}\right)\right).$$

Adding the contribution from $\mathcal{J}_2$ and replacing $c_1 \to c$ and $c_2 \to \bar{c}$, the total joint contribution to the subregion action is

$$S_{\mathcal{J}}^R = 2L^2\Delta x\left(\frac{1}{z_H}\log\left(\frac{c\bar{c}\, z_H^2(z_H + \delta_R)}{4L^2(z_H - \delta_R)}\right) - \frac{1}{\delta_R}\log\left(\frac{c\bar{c}\,\delta_R^2 z_H^2}{L^2(z_H^2 - \delta_R^2)}\right)\right).$$

The counterterm contributions from null boundaries $\mathcal{N}_3$ and $\mathcal{N}_4$ vanish and the remaining pieces give

$$\Delta S_{\mathcal{N}_1}^R + \Delta S_{\mathcal{N}_2}^R = 4L^2\Delta x\left(\frac{1}{\delta_R} - \frac{1}{z_H}\right) + 2L^2\Delta x\left(\frac{1}{\delta_R}\log\left(\frac{c\bar{c}\,\tilde{L}_1\tilde{L}_2\delta_R^2}{L^4}\right) - \frac{1}{z_H}\log\left(\frac{c\bar{c}\,\tilde{L}_1\tilde{L}_2 z_H^2}{L^4}\right)\right).$$

Combining the previous expressions we find the total result for the subregion action without ($\mathcal{A}^R$) and with ($\tilde{\mathcal{A}}^R$) the counterterms:

$$\mathcal{A}^R = \frac{L^2\Delta x}{8\pi G}\left(\frac{1}{z_H}\left(2 + \log\left(\frac{c\bar{c}\, z_H^2(z_H + \delta_R)}{4L^2(z_H - \delta_R)}\right)\right) - \frac{1}{\delta_R}\left(2 + \log\left(\frac{c\bar{c}\,\delta_R^2 z_H^2}{L^2(z_H^2 - \delta_R^2)}\right)\right)\right),$$

$$\tilde{\mathcal{A}}^R = \frac{L^2\Delta x}{8\pi G}\left(\frac{1}{z_H}\log\left(\frac{L^2(z_H + \delta_R)}{4\tilde{L}_1\tilde{L}_2(z_H - \delta_R)}\right) - \frac{1}{\delta_R}\log\left(\frac{L^2 z_H^2}{\tilde{L}_1\tilde{L}_2(z_H^2 - \delta_R^2)}\right)\right). \tag{A8}$$

In the main text, our discussion centers on the result with counterterms included (reported there in a somewhat different form, see Sec. V B).